\documentclass[aps,prb,floatfix,amsmath,amssymb,twocolumn, 10pt]{revtex4-1}
\usepackage[american]{babel}
\usepackage{graphicx}
\usepackage{dcolumn}
\usepackage{bm, epsfig}
\usepackage{hyphenat}
\usepackage{amssymb}   
\usepackage{verbatim}

\setcitestyle{super}

\newcommand*{\citen}[1]{%
  \begingroup
    \romannumeral-`\x 
    \setcitestyle{numbers}%
    \cite{#1}%
  \endgroup   
}

\newcommand{\beginsupplement}{%
        \setcounter{table}{0}
        \renewcommand{\thetable}{S\arabic{table}}%
        \setcounter{figure}{0}
        \renewcommand{\thefigure}{S\arabic{figure}}%
        \setcounter{section}{0}
        \renewcommand\thesection{\Alph{section}}
        \setcounter{subsection}{0}
	\renewcommand\thesubsection{\thesection\arabic{subsection}}
        \setcounter{equation}{0}
	\renewcommand\theequation{S\arabic{equation}}
     }

\begin{document}

\title{Narrow-band hard-x-ray lasing}


\author{Chunhai Lyu}
\affiliation{Max-Planck-Institut f\"{u}r Kernphysik, Saupfercheckweg 1, 69117 Heidelberg, Germany}
\author{Stefano M. Cavaletto}
\email[Corresponding author. Email: ]{smcavaletto@gmail.com}
\affiliation{Max-Planck-Institut f\"{u}r Kernphysik, Saupfercheckweg 1, 69117 Heidelberg, Germany}
\author{Christoph H. Keitel}
\affiliation{Max-Planck-Institut f\"{u}r Kernphysik, Saupfercheckweg 1, 69117 Heidelberg, Germany}
\author{Zolt\'{a}n Harman}
\affiliation{Max-Planck-Institut f\"{u}r Kernphysik, Saupfercheckweg 1, 69117 Heidelberg, Germany}

\date{\today}

\maketitle

\textbf{Since the advent of x-ray free-electron lasers (XFELs), considerable efforts have been devoted to achieve x-ray pulses 
with better temporal coherence~\cite{lambert2008injection,Amann2012demonstration,Allaria2012highly,Ratner2015Experimental,Huang2017Generating,Rohringer2012,Yoneda2015atomic,Kim2008AProposal}. 
Here, we put forward a scheme to generate fully coherent x-ray lasers (XRLs) 
based on population inversion in highly charged ions (HCIs),
created by fast inner-shell photoionization using XFEL pulses in a laser-produced plasma.
Numerical simulations show that one can obtain high-intensity, femtosecond x-ray pulses of relative bandwidths $\Delta\omega/\omega=10^{-5}$ -- $10^{-7}$
by orders of magnitude narrower than in XFEL pulses for wavelengths down to the sub-\aa{ngstr\"{o}m}~regime. 
Such XRLs may be applicable in the study of x-ray quantum optics~\cite{adams2013x,Vagizov2014coherent,heeg2017spectral,Haber2017Rabi} and metrology~\cite{Cavaletto2014Broadband},
investigating nonlinear interactions between x-rays and matter~\cite{Kanter2011,Doumy2011}, or in
high-precision spectroscopy studies in laboratory astrophysics~\cite{Bernitt2012}.} 

Most of the XFEL facilities in operation or under construction generate x-ray pulses based on the self-amplified spontaneous-emission (SASE) process. 
Despite their broad ($\Delta\omega/\omega\sim10^{-3}$) and chaotic spectrum~\cite{Pellegrini2016}, such high-intensity SASE x-ray pulses have found diverse applications 
in physics, chemistry and biology~\cite{Bostedt2016,Seddon2017short}. 
In order to improve the temporal coherence and frequency stability, 
different seeding schemes have been implemented successfully~\cite{lambert2008injection,Amann2012demonstration,Allaria2012highly,Ratner2015Experimental,Huang2017Generating}. 
In the hard-x-ray regime, the self-seeding mechanism has reduced the relative bandwidth to the level of $5\times10^{-5}$ at photon energies of 8~--~9~keV~(ref.~\citen{Amann2012demonstration}). 
However, at higher energies around 30~keV, the predicted relative bandwidth for seeded XFELs is still around $4\times10^{-4}$~(ref.~\citen{Pellegrini2016}). 
Further reduction of the bandwidth with low-gain XFEL oscillators (XFELOs) has also been proposed. 
By recirculating the x-ray pulses through an undulator in a cavity, the output x-rays have an estimated relative bandwidth as small as $10^{-7}$ (ref.~\citen{Kim2008AProposal}). 
To date, however, the XFELO scheme remains untested. 

\begin{figure}[b]
\renewcommand{\figurename}{Figure}
\renewcommand{\thefigure}{\arabic{figure}}
\includegraphics[width=0.45\textwidth]{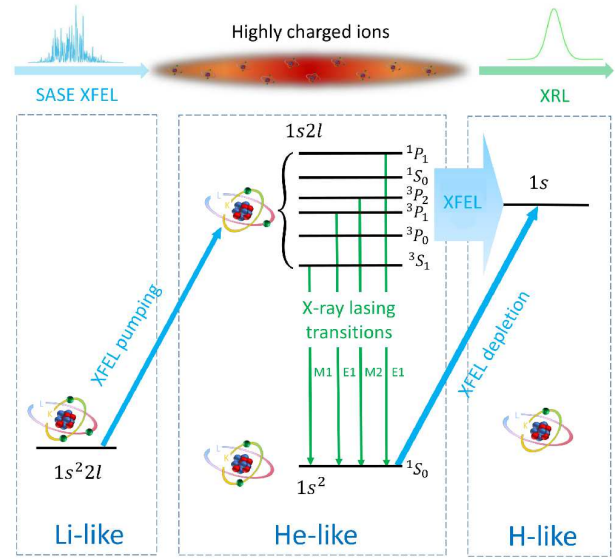}
\caption{\label{fig:LasingScheme} \textbf{Scheme of the lasing process.} An upper lasing state $1s2l$
of He-like ions is pumped through K-shell photoionization of Li-like ions initially in a $1s^22l$ state by an XFEL pulse tuned above the K-edge of
the He-like ions (blue arrow left).
Lasing takes place through one of the four possible transitions from an upper lasing state $1s2l$ to the lower lasing state $1s^2$ (green arrows). The state $1s^2$ is
depleted through further K-shell photoionization by the same XFEL pulse (blue arrow right). L-shell photoionization of the upper lasing states is represented
by the thick blue arrow. The order of the $1s2l$ states may vary for different elements.}
\end{figure}

Plasma-based XRLs have achieved saturated amplification for different wavelengths in the soft-x-ray regime. 
Such lasers are mainly based on the $3p\rightarrow3s$ or $4d\rightarrow4p$ transition in Ne- or Ni-like ions for elements varying from $^{14}$Si to $^{79}$Au, 
where the population inversion is achieved through electron collisional excitation in a hot dense plasma~\cite{Daido2002review,Suckewer2009x}. 
The first hard-XRL was initially proposed through direct pumping of the $1s^{-1}\rightarrow2s^{-1}$ transition via photoionization of a K-shell electron~\cite{duguay1967some}. 
Limited by the high pump power required, this scheme was demonstrated only in recent years after high-flux XFELs became available~\cite{Rohringer2012,Yoneda2015atomic}. 
In the experiments with solid copper, an XRL at 1.54~\AA ~was obtained, with a measured relative bandwidth of $2\times10^{-4}$ determined by the 
short lifetime of the upper lasing state due to fast Auger decay. Here we put forward an XFEL-pumped transient XRL based on He-like HCIs, 
where the detrimental Auger-decay channel is nonexistent 
due to the lack of outer-shell electrons. 
By choosing a lasing transition which slowly decays radiatively, this results in a further reduction of the XRL bandwidth by several orders of magnitude. 

\begin{table*}[t]
\renewcommand{\tablename}{Table}
\renewcommand{\thetable}{\arabic{table}}
\caption{\textbf{Results for selected x-ray lasing transitions for different elements.} 
Radiative parameters like transition energy $ \omega_{\text{0}}$, natural linewidth $\Gamma $ and upper-state lifetime $\tau$
are obtained from the GRASP code~\cite{Dyall1989grasp}.  
The XFEL photon energy $\omega_{\text{xfel}}$, tuned above the K-shell ionization threshold of the $1s^2$ state, 
and the mean peak flux (ph.~stands for photons), used in the 1,000 realizations of the SASE XFEL pulses, are fixed to ensure population inversion and the applicability of the two-level approximation. 
Electron temperature $T_{\text{e}}$ and ion density $N_{\text{i}}$ ($n_{\text{0}}=10^{20}\text{ cm}^{-3}$) are chosen to enable 
a significant fraction of Li-like ions in the plasma~\cite{Chung2005flychk} (Supplementary Fig.~S1). 
The broadening effects, with Doppler broadening $\Delta\omega_{\text{D}}$, electron--ion impact broadening $\Delta\omega_{\text{e-i}}$,
and ion--ion Stark broadening $\Delta\omega_{\text{i-i}}$, are calculated for given $T_{\text{e}}$ and $N_{\text{i}}$ based on 
Maxwell--Boltzmann distributions~\cite{Griem1974}, with ion temperature $T_{\text{i}}=487$~K. 
The characteristic length $L_{\text{c}}$ is the optimal length defined in Fig.~\ref{fig:peakinstensity_pulsedurantion_spectrumwidth}. 
The XRL intensity $I_{\text{c}}$ as well as the relative bandwidth $\Delta\omega/\omega_0$ at this length are obtained by averaging over 1,000 simulations, 
with the uncertainties arising from the random XFEL pulse profile. The upper and lower bounds are 
the values at the 10th and 90th percentiles of the corresponding distributions 
(as shown in Figs.~\ref{fig:peakinstensity_pulsedurantion_spectrumwidth}d,f for $\text{Ar}^{\text{16+}}~^3P_1  $).}
\begin{ruledtabular}
\begin{tabular}{cccc|cc|ccccc|ccc}
  \multicolumn{4}{c}{Radiative parameters} & \multicolumn{2}{c}{XFEL} & \multicolumn{5}{c}{Plasma conditions} & \multicolumn{3}{c}{Simulation results}\\
Upper state     &     $ \omega_{\text{0}}$ & $\Gamma $     & $\tau$ &   $ \omega_{\text{xfel}}$  &     peak flux & $T_{\text{e}}$ & $N_{\text{i}}$ & $\Delta\omega_{\text{D}}$ & $\Delta\omega_{\text{e-i}}$  & $\Delta\omega_{\text{i-i}}$ & $L_{\text{c}}$& $I_{\text{c}}$  & $\Delta\omega/\omega_0$\\
$1s2l$           &       $     \text{ (keV)}$ & $\text{ (meV)}$  &   (ps) &   (keV) &       (ph./cm$^2$/s) &  (keV) &  ($n_{\text{0}}$)    & (meV)  & (meV) & (meV) & (mm) &$\text{(W}/\text{cm}^{2}\text{)}$ &  \\
\hline
$\text{Ne}^{\text{8+}}~^1P_1   $ &   $0.920$   & $6.08$ & $0.10$  & 1.197  &   $1.10\times10^{34}$ & 0.035  & $0.02$  & 3.23  & 0.08  & 0.06& 2.8 &$4.5\substack{_{+9.3} \\ ^{-4.4}}\times10^{12} $ & $7.8\substack{^{+4.3} \\ ^{-2.7}}\times10^{-5}$\\
$\text{Ar}^{\text{16+}}~^3P_1  $ &   $3.120$   & $1.16$ & $0.57$  & 4.124  &  $2.29\times10^{33}$  & 0.25  & $0.25 $ & 7.80  & 1.38  & 9.81  & 3.3 &$5.0\substack{_{+3.8} \\ ^{-4.0}}\times10^{14} $ & $7.8\substack{^{+1.9} \\ ^{-2.1}}\times10^{-6}$ \\
$\text{Ar}^{\text{16+}}~^1P_1  $ &   $3.137$   & $72.2$ & $0.009$ & 4.124  &   $1.98\times10^{35}$  & 0.25  & $2.48 $ & 7.83  & 0.29  & 25.22& 0.35  &$1.5\substack{_{+1.4} \\ ^{-0.8}}\times10^{16} $  & $4.7\substack{^{+2.0} \\ ^{-1.7}}\times10^{-5}$\\
$\text{Kr}^{\text{34+}}~^3P_2  $ &   $13.087$  & $0.07$ & $9.46$  & 17.316 &   $1.87\times10^{34}$ &  6.20  & $ 0.50 $  & 22.58 & 0.45   & 2.76& 289  &$4.3\substack{_{+7.6} \\ ^{-4.2}}\times10^{16} $  & $3.7\substack{^{+1.6} \\ ^{-1.1}}\times10^{-7}$\\
$\text{Xe}^{\text{52+}}~^3P_2  $ &   $30.589$  & $1.91$ & $0.34$  & 40.304 &   $7.45\times10^{34}$ & 25.00  & $ 27.0 $ & 42.16 & 7.60   & 126& 8.5  &$4.7\substack{_{+6.9} \\ ^{-4.5}}\times10^{18} $  & $1.5\substack{^{+0.6} \\ ^{-0.5}}\times10^{-6}$\\
\end{tabular}
\label{Transitions}
\end{ruledtabular}
\end{table*}

The photoionization-pumped atomic laser is illustrated in Fig.~\ref{fig:LasingScheme}, where the Li-like HCIs are initially prepared in a $1s^22l$ ($l=s,p$) state in a 
laser-produced plasma~\cite{Chung2005flychk}. 
A SASE XFEL pulse tuned above the K-edge of the ions first removes a K-shell electron from the Li-like ions, creating He-like ions in the $1s2l$ excited states.
Subsequent decay to the $1s^2$ ground state leads to emission of x-ray photons via four possible K$\alpha$ transitions:
one magnetic-dipole ($M1$) transition from the $^3S_1$ state,
two electric-dipole ($E1$) transitions from the $^3P_1$ or $^1P_1$ states, and one magnetic-quadrupole ($M2$) transition from the $^3P_2$ state.
The population inversion in the He-like ions resulting from this photoionization-pumping scheme leads to amplification of the emitted x-rays, 
i.e., to inner-shell x-ray lasing.

Two factors determine which transition will lase. 
Firstly, population inversion is needed to have stimulated emission.
Typical XFEL facilities, with
peak photon fluxes of $10^{33}$ -- $10^{35}\text{~cm}^{-2}\text{~s}^{-1}$ (1~$\mu$m$^2$ spot size, ref.~\citen{Pellegrini2016,Seddon2017short}),
yield an inverse ionization rate of a few femtoseconds, such that XFEL pulses can effectively photoionize all the Li-like ions.
Transitions with upper-state lifetimes longer than 1~fs are necessary to ensure population inversion
between the $1s2l$ and $1s^2$ state. 
Furthermore, the finite lifetime of the plasma $\tau_{\text{p}}\sim10$~ps~(ref.~\citen{KRAINOV2002237}) also influences the lasing process. Sufficient amplification of x-ray
radiation will take place only from transitions whose decay is slower than XFEL pumping, but faster than plasma expansion.

Suitable XRL transitions in He-like ions are shown in table~\ref{Transitions}.
Simulations for each transition have been conducted by solving the Maxwell--Bloch equations~\cite{Larroche2000maxwell,Weninger2014} 
(Supplementary Information) numerically 
in retarded-time coordinates for 
1,000 different realizations of SASE XFEL pulses. 
For most of the ions displayed in table~\ref{Transitions}, only one of the four transitions in Fig.~\ref{fig:LasingScheme} satisfies the requirements for lasing described above, 
such that a two-level description of the He-like ions is applicable. 
For $\text{Ar}^{\text{16+}}$ ions, there are two transitions which may lase simultaneously. 
Thus, the XFEL peak flux is tuned properly to ensure that only one of them lases. Values of the initial populations of the states in Li- and He-like ions are 
computed with the FLYCHK code~\cite{Chung2005flychk}. 
XFEL frequencies and photon fluxes are then fixed such that K-shell photoionization of the $1s^22l$ and $1s^2$ states ensures population inversion 
in the He-like ions (Supplementary Information). 
Except for the transition from $\text{Kr}^{\text{34+}}~^3P_2 $, which needs
more than 10~cm to reach saturated intensity, all the other transitions are predicted to generate high-intensity x-ray pulses within 1 cm with small bandwidths. 
For $E1$ transitions in $\text{Ne}^{\text{8+}}$ and $\text{Ar}^{\text{16+}}$, a significant improvement of $\Delta\omega/\omega$ is obtained compared to 
SASE XFEL pulses~\cite{Pellegrini2016} and XRLs with neutral atoms~\cite{Rohringer2012,Yoneda2015atomic}. 
When going to heavier $\text{Kr}^{\text{34+}}$ and $\text{Xe}^{\text{52+}}$ ions, the $M2$ transitions provide an even more significant reduction of the bandwidth, 
with $\Delta\omega/\omega_0$ being $3.7\times10^{-7}$ and $1.5\times10^{-6}$, respectively. 
The resulting $13$- and $30$-keV lasers feature similar bandwidths as the untested XFELO scheme~\cite{Kim2008AProposal}, with intensities of $\sim10^{18}\text{~W~cm}^{-2}$. 
The relative bandwidths are by 2 to 3 orders of magnitude narrower than the value predicted for future seeded-XFEL sources at analogous hard-x-ray 
wavelengths around 0.41~--~0.95~\AA~(ref.~\citen{Pellegrini2016}).

\begin{figure*}[t]
\renewcommand{\figurename}{Figure}
\renewcommand{\thefigure}{\arabic{figure}}
\includegraphics[width=0.9\textwidth]{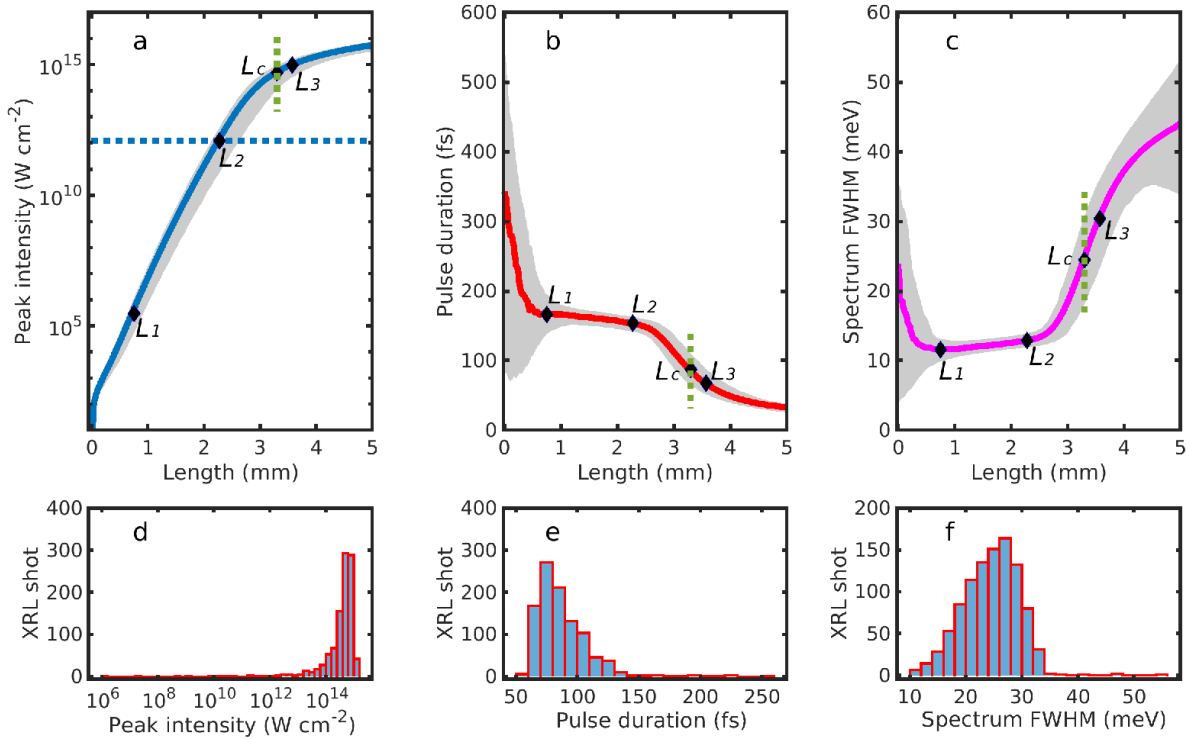}
\caption{\label{fig:peakinstensity_pulsedurantion_spectrumwidth} \textbf{Evolution of the XRLs over 1,000 simulations ($\text{Ar}^{\text{16+}}~^3P_1$)}. \textbf{a}-\textbf{c}, Peak intensity, pulse duration 
and spectral full width at half maximum (FWHM) for the x-ray laser. The solid lines display results averaged over 1,000 simulations. 
The dotted line in (\textbf{a}) indicates the saturation intensity $I_{\text{s}}=1.18\times10^{12}\text{~W~cm}^{-2}$. 
$L_1$, $L_2$ and $L_3$ mark the lengths for the XRL pulse to reach transform-limited profile, saturation intensity and Rabi flopping, respectively. $L_\text{c}$ refers to the 
characteristic length that optimizes the intensity and bandwidth of the XRL pulses. 
Here, it is defined as the length at which the slope of the solid line in (\textbf{a}) is 1/3 of the slope at $L_2$. 
The gray areas in (\textbf{a-c}) indicate the distribution areas of the results over 1,000 simulations. At a given length, 
the bottom and top edges of the areas indicate the 10th and 90th percentiles of the distributions, respectively. 
\textbf{d-f}, Distributions of the peak intensity, pulse duration and spectral FWHM at $L_{\text{c}}$ along the green dotted lines in 
(\textbf{a-c}). 
In (\textbf{d}), there are 15 simulations whose peak intensities locate in the unsaturated region $10^6\sim10^{12}\text{~W~cm}^{-2}$, indicating that 1.5\% of the 
SASE XFEL pulses cannot provide enough pumping.}
\end{figure*}

\begin{figure*}[t]
\renewcommand{\figurename}{Figure}
\renewcommand{\thefigure}{\arabic{figure}}
\includegraphics[width=0.9\textwidth]{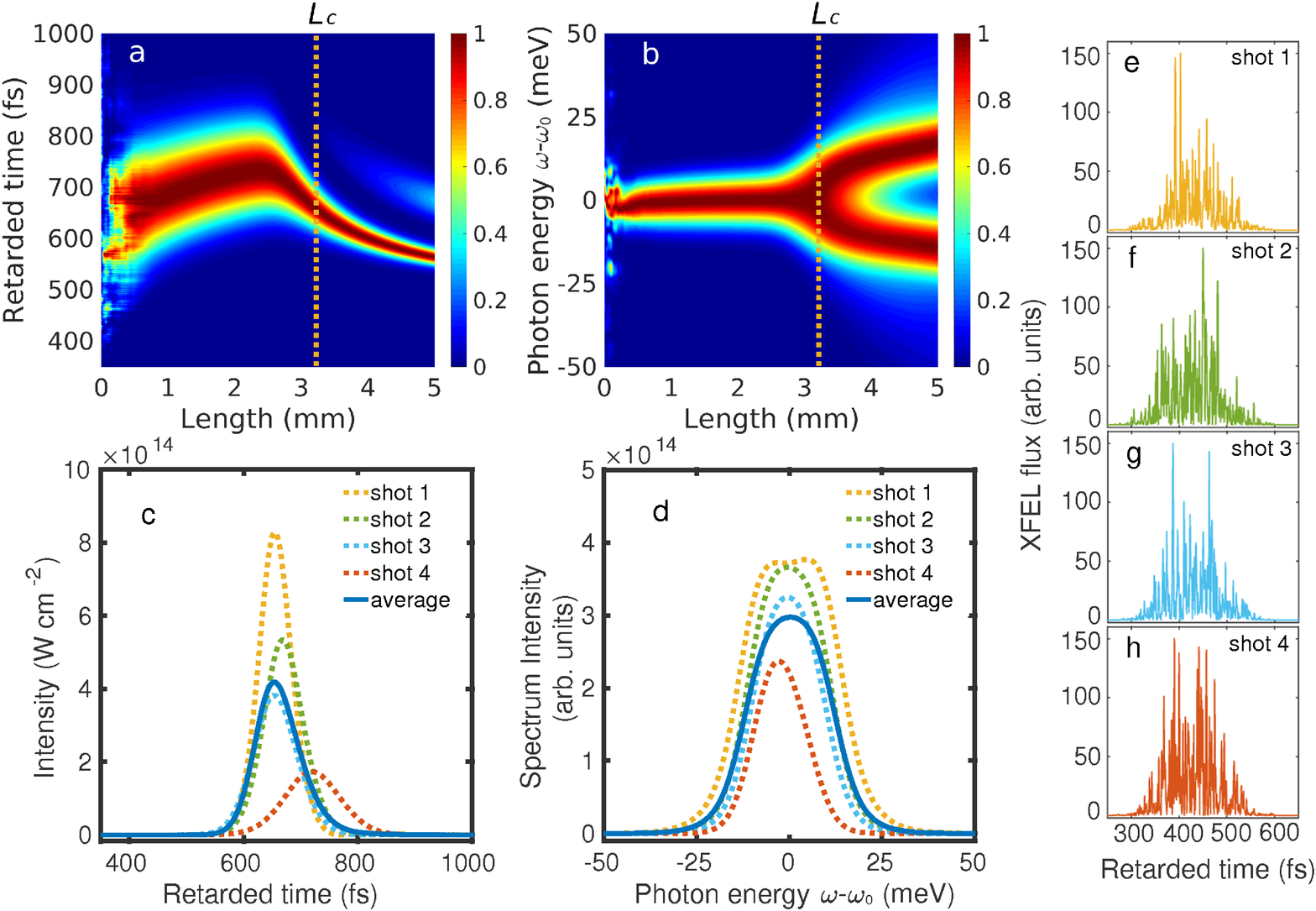}
\caption{\label{fig:2D_Normalized_Intensity_Spectrum} \textbf{Evolution of the normalized XRL intensity and spectrum ($\text{Ar}^{\text{16+}}~^3P_1$, single simulation).} 
\textbf{a}, Intensity shown as a function of retarded time and propagation length. 
\textbf{b}, Power spectrum displayed as a function of photon energy and propagation length. For a given length, the intensity and spectrum are normalized to the maximum value of the corresponding 
profiles at that length. 
The vertical dotted lines indicate the characteristic length $L_{\text{c}}$ shown in Fig.~\ref{fig:peakinstensity_pulsedurantion_spectrumwidth}. 
For lengths larger than $L_{\text{c}}$, the strength of the second peak in (\textbf{a}) appearing around retarded times of $\sim700$~fs 
has been multiplied by a factor of~5 for better visibility. 
\textbf{c},\textbf{d}, XRL pulse profile and spectrum at $L_{\text{c}}$. The yellow dotted lines correspond to the results from the simulation in (\textbf{a},\textbf{b}). 
Three other simulation results (green, blue and red dotted lines) at $L_{\text{c}}$ are also included for comparison, 
with the solid lines corresponding to the results averaged over 1,000 simulations. 
\textbf{e-h}, SASE XFEL pulses used in the four simulations in (\textbf{c},\textbf{d}).
Differences in amplitudes and positions of the peak intensities 
in (\textbf{c},\textbf{d}) are the result of shot-to-shot random profiles of the SASE XFEL pulses. 
XRL intensity and spectrum as a function of propagation length for the XFEL pulses in (\textbf{f-h}), and averaged over 1,000 simulations, can be found in 
the Supplementary Information. 
In (\textbf{c},\textbf{d}), the blue solid lines are obtained by averaging XRL pulse shapes and spectra as a function of retarded time and frequency, respectively. 
The peak intensity of the averaged XRL pulse shape in (\textbf{c}), 
therefore, differs from the averaged value of the XRL peak intensity displayed in Fig.~\ref{fig:peakinstensity_pulsedurantion_spectrumwidth}a and table~\ref{Transitions}, 
because the position of the intensity peak varies from shot to shot.}
\end{figure*}

To understand the properties of our XRLs and how they develop in the plasma, simulation results for the 
$^3P_1\rightarrow~^1S_0$ transition in Ar$^{16+}$ are shown in Figs.~\ref{fig:peakinstensity_pulsedurantion_spectrumwidth} and~\ref{fig:2D_Normalized_Intensity_Spectrum} 
for the corresponding parameters listed in table~\ref{Transitions}.
We use a partial-coherence method~\cite{Cavaletto2012Resonance} to simulate $124$-fs-long SASE XFEL pulses with a spectral width of $1.55$~eV and 
peak photon flux of $2.29\times10^{33} \text{~cm}^{-2}\text{~s}^{-1}$. 
This results in a peak pumping rate of $6.04\times10^{13} \text{~s}^{-1}$ for the upper lasing state, 
and a depletion rate of $5.63\times10^{13} \text{~s}^{-1}$ for the lower lasing state. 
They are $32$ and $58$ times larger than the spontaneous-emission rate of the $^3P_1$ state, ensuring population inversion. 

Average results over 1,000 SASE-pulse realizations are shown in Figs.~\ref{fig:peakinstensity_pulsedurantion_spectrumwidth}a-c. 
The peak intensity of the XRL, shown by the solid line in Fig.~\ref{fig:peakinstensity_pulsedurantion_spectrumwidth}a, increases exponentially during the initial propagation stage, 
then displays a saturation behavior. 
The dotted line indicates the saturation intensity $I_{\text{s}}=\hbar\Gamma\omega_0^3/6\pi c^2$  
at which the stimulated-emission rate equals the spontaneous-emission rate. This is also the intensity from which the amplification begins to slow down. 
The evolution of the pulse duration and the spectral width are shown by the solid lines 
in Figs.~\ref{fig:peakinstensity_pulsedurantion_spectrumwidth}b,c. 
At $L=0$, only spontaneous emission takes place: the 342-fs average pulse duration is mainly determined by the lifetime of the $^3P_1$ state 
(Fig.~\ref{fig:peakinstensity_pulsedurantion_spectrumwidth}b), 
whereas the 23.5-meV intrinsic spectral width before propagation (Fig.~\ref{fig:peakinstensity_pulsedurantion_spectrumwidth}c) is mostly due to the sum of the natural linewidth~$\Gamma$ 
and the three broadening effects shown in table~\ref{Transitions}.

During its propagation in the medium, gain narrowing and saturation rebroadening will also contribute to the final bandwidth. 
This can be observed by inspecting the four distinct propagation regions separated by $L_1$, $L_2$ and $L_3$ in Figs.~\ref{fig:peakinstensity_pulsedurantion_spectrumwidth}a-c, 
which can also be followed in Figs.~\ref{fig:2D_Normalized_Intensity_Spectrum}a,b for a single simulation. 
Up to $L_1=0.75$~mm, both the pulse duration and spectral FWHM decrease severely. 
The laser intensity and spectrum in this region for a single simulation are spiky and noisy, 
as the ions irradiate randomly in time and space (Figs.~\ref{fig:2D_Normalized_Intensity_Spectrum}a,b). 
When the spontaneously emitted signal propagates and stimulated emission sets in, it selectively amplifies the frequencies 
around $\omega_0$ such that the XRL pulse approaches a fully coherent transform-limited profile at $L_1$~(ref.~\citen{Weninger2014}), 
with a bandwidth smaller than the intrinsic width. 
Thereafter, a gradual broadening of the spectrum is observed in the region $L_1L_2$. 
The broadening increases abruptly from $L_2=2.3$~mm, where the saturation intensity has been reached and the stimulated-emission rate exceeds the spontaneous-emission rate. 
This is accompanied by a substantial slowing down of the amplification of the intensity 
and a significant decrease of the pulse duration in the region between $L_2$ and $L_3$ (Figs.~\ref{fig:peakinstensity_pulsedurantion_spectrumwidth}a,b). 
Further propagation of the XRL pulse after $L_3=3.5$~mm is characterized by the onset of Rabi flopping (Fig.~\ref{fig:2D_Normalized_Intensity_Spectrum}a) 
which is reflected by a splitting in the XRL spectrum (Fig.~\ref{fig:2D_Normalized_Intensity_Spectrum}b). 
This effect is much stronger and more marked than for previous XFEL-pumped transient lasers with neutral atoms (ref.~\citen{Weninger2014}) 
due to the absence of Auger decay~\cite{Cavaletto2012Resonance}.
At the same time, the gain of the laser intensity in this region is strongly suppressed. 

The optimal choice for a coherent XRL pulse is located in the third region $L_2L_3$, where saturation has already been reached while the bandwidth is still narrow. 
By choosing the medium length to be $L_{\text{c}}=3.3$~mm, as shown in Fig.~\ref{fig:peakinstensity_pulsedurantion_spectrumwidth}, 
one will obtain an approximately $87$-fs-long XRL pulse with an average peak intensity of 
$I_{\text{c}}=5.0\times10^{14}\text{~W~cm}^{-2}$ ($\sim$80\% fluctuations) and an average bandwidth of $\Delta\omega=24.5$~meV ($\sim$30\% fluctuations) 
(table~\ref{Transitions}). 
This gives $\Delta\omega/\omega_0=7.8\times10^{-6}$ for a total of $6.5\times10^{8}$ coherent photons, 
with a peak brilliance of $1.4\times10^{31}$ photons/s/mm$^2$/mrad$^2$/0.1\%bandwidth.

In conclusion, we put forward a scheme to obtain high-intensity x-ray lasers with 
bandwidths up to three orders of magnitude narrower compared to the value predicted for seeded-XFEL sources in the hard-x-ray regime~\cite{Pellegrini2016}. 
While the photon energies of the XRLs are fixed by the corresponding isolated transitions, three-wave interactions by 
synchronizing XFEL pulses with optical/XUV lasers could be investigated in future studies to tune the frequencies in a broad range. 
Such x-ray sources will enable novel studies of coherent light--matter interactions in atomic, molecular and solid-state systems.


\noindent The authors thank S.~Tang, Y.~Wu and J.~Gunst for fruitful discussions. C.L.~developed the mathematical model, performed the analytical calculations and numerical simulations, and wrote the manuscript. 
S.M.C.~and Z.H.~conceived the underlying scheme. All authors contributed to the development of ideas, 
discussion of the technical aspects and results, and preparation of the manuscript. 
Correspondence and requests for materials should be addressed to S.M.C.~(smcavaletto@gmail.com).\\

%
%
%
%
\newpage

\onecolumngrid
\setcounter{section}{0}
\section*{Supplementary Information}
\beginsupplement

\section{Modeling of the propagation through the medium} \label{A}
In this section, we first discuss the model used to describe lasing in He-like ions in the two-level approximation. 
Maxwell--Bloch equations are developed to model the dynamics of the system and the propagation of the x-ray laser (XRL) pulse through the gain medium. 
The equations are solved numerically with a Runge--Kutta method in the retarded time domain. 
Initial populations of the ionic states are obtained from FLYCHK simulations of the charge-state distributions under given plasma conditions. 
Cross sections of XFEL photoionization  for each element, as well as XFEL pulse durations and bandwidths used in the simulations, are also listed.

\subsection{Maxwell--Bloch equations in the two-level approximation} 
\label{A1}
Selected transitions applicable for our lasing scheme are shown in table~1 from the main text. 
A full description of the lasing process should account for all the $1s2l$ ($l=s,p$) states in He-like ions. 
However, the lifetimes of these states differ from each other by orders of magnitude. 
When only one of the four K$\alpha$ transitions in the He-like ions satisfies the lasing requirements described in the main text, 
a two-level description of the ions is sufficient.

For light ions, Ne$^{8+}$ for example, the $E1$ transition with a decay rate of $9.24\times10^{12}\text{ s}^{-1}$ from the $^1P_1$ state will develop lasing. 
The other transitions have decay times much larger than the plasma expansion time, and their contribution is negligible compared to the $^1P_1$ state, such that they can be neglected. 
For heavy ions like Xe$^{52+}$, however, the two $E1$ transition rates scale as $\sim Z^4$ ($Z$ being the atomic number), 
corresponding to $3.05 \times10^{15}\text{ s}^{-1}$ for $^3P_1$ and $6.82 \times10^{15}\text{ s}^{-1}$ for $^1P_1$, respectively,
which are too large to enable population inversion with available XFEL pulses. 
On the other hand, the decay rate of the $M1$ transition from the $^3S_1$ state and the $M2$ transition from the $^3P_2$ state 
are $3.7\times10^{11}\text{ s}^{-1}$ and $2.56\times10^{12}\text{ s}^{-1}$, respectively, 
which are sufficient for lasing to take place before the expansion of the plasma. 
However, the $M1$ transition from the $^3S_1$ state is dominated by large Stark broadening effects, such that 
the amplification of photons emitted from such transition is much slower than for those emitted in the $M2$ transition. 
Within the characteristic length $L_{\text{c}}$ for the $M2$ transition, the presence of the $^3S_1$ state can hence be neglected. 
Similar arguments are applicable to Kr$^{34+}$. 

For $\text{Ar}^{\text{16+}}$, 
there are two transitions that may lase simultaneously, as listed in table~1 from the main text. 
However, the lifetimes $\tau$ of these states differ from each other by a factor of 63. 
The XFEL photon flux can be tuned properly to exclude one of them from lasing. 
For instance, in order to obtain lasing from the $^3P_1\rightarrow~^1S_0$ transition, a mean peak flux of $2.29\times10^{33} \text{ cm}^{-2}\text{ s}^{-1}$ for the XFEL pulses is applied. 
This generates an upper state pumping rate and a lower-state depletion rate which are $32$ and $58$ times larger than the decay rate of the $^3P_1$ upper state, respectively, 
resulting in population inversion of the transition. 
At the same time, for this value of the peak flux, the pumping rate of the $^1P_1$ upper state is only 0.26 times the decay rate of the $^1P_1$ state, too small to obtain population inversion 
in the  $^1P_1\rightarrow~^1S_0$  transition. Thereby, lasing will only take place from the $^3P_1$ state. 
Sufficient amplification of the x-ray photons emitted from the $^1P_1$ state needs higher XFEL peak fluxes and higher ion densities. 
When such conditions are met, e.g., for the parameters shown in table~1 in the main text, lasing from the $^1P_1\rightarrow~^1S_0$ transition takes place. 
Saturation will be reached much sooner than for the $^3P_1\rightarrow~^1S_0$ transition, such that the $^3P_1$ state can be neglected.

Assuming XFEL pulses propagating along the $\hat{x}$ direction, 
the evolution of the x-ray laser field in the slowly varying envelope approximation is given by~\cite{Larroche2000maxwellS,Weninger2014S}
\begin{eqnarray}
\frac{\partial\mathcal{A}(x,t)}{\partial t}+c\frac{\partial\mathcal{A}(x,t)}{\partial x}=
i\frac{\mu_{\text{0}}\omega_{\text{0}}c^2}{2}\mathcal{F}(x,t),
\label{MexwellEquation}
\end{eqnarray}
where $\mathcal{A}(x,t)$ is either the electric field $\mathcal{E}(x,t)$ or the magnetic field $\mathcal{B}(x,t)$, depending on the specific transition. 
$\mu_{\text{0}}$ is the vacuum permeability and $c$ is the vacuum speed of light.
$\mathcal{F}(x,t)$ corresponds to the polarization field induced by $\mathcal{E}(x,t)$ for $E1$ transitions, or the magnetization field 
and magnetic-quadrupole field induced by $\mathcal{B}(x,t)$ for $M1$ transitions and $M2$ transitions, respectively. 

For a given lasing transition, we assume all Li-like ions are pumped into the corresponding upper lasing state of such transition by the XFEL pulse.
Using $\left|\text{e}\right>$ and $\left|\text{g}\right>$ to represent the upper lasing state and the lower lasing state, respectively,
the dynamics of the He-like ions are described by the Bloch equations of the density matrix
\begin{eqnarray}
\dot{\rho}_{\text{ee}}(x,t)&=&-\text{Im[}\Omega^*(x,t)\rho_{\text{eg}}(x,t)\text{]}+\sigma_{\text{0}}j_{\text{xfel}}(x,t)\rho_{\text{00}}(x,t) 
\label{drhoee}
-\sigma_{\text{e}}j_{\text{xfel}}(x,t)\rho_{\text{ee}}(x,t)-\Gamma\rho_{\text{ee}}(x,t), \\
\dot{\rho}_{\text{eg}}(x,t)&=&\frac{i}{2}\Omega^*(x,t)(\rho_{\text{ee}}(x,t)-\rho_{\text{gg}}(x,t))-\frac{\gamma}{2}\rho_{\text{eg}}(x,t) 
\label{drhoeg}
+S(x,t),\\
\dot{\rho}_{\text{gg}}(x,t)&=&\text{Im[}\Omega^*(x,t)\rho_{\text{eg}}(x,t)\text{]}-\sigma_{\text{g}}j_{\text{xfel}}(x,t)\rho_{\text{gg}}(x,t) 
+\Gamma\rho_{\text{ee}}(x,t).
\label{drhogg}
\end{eqnarray}
The carrier frequency $\omega_{\text{0}}$ is chosen to be resonant with the lasing transition.
$\rho_{\text{ee}}$ and $\rho_{\text{gg}}$ are the populations
of $\left|\text{e}\right>$ and $\left|\text{g}\right>$, and the off-diagonal term $\rho_{\text{eg}}$ represents the coherence between the two lasing states.
$\Omega(x,t)=\wp \mathcal{E}(x,t)/\hbar$ for an $E1$ transition (or $\Omega(x,t)=m \mathcal{B}(x,t)/\hbar$ for an $M1$ transition,
and $\Omega(x,t)=k_{0}q_{yx} \mathcal{B}(x,t)/\hbar$ for an $M2$ transition \cite{PhysRevC.77.044602S})
is the time- and space-dependent Rabi frequency,
with $\wp$ the electric-dipole moment, $m$ the magnetic-dipole moment, and $q_{yx}$ the $yx$-component of the magnetic-quadrupole tensor.
XFEL pumping of the $\left|\text{e}\right>$ state from Li-like ions is accounted for through the second term in the right-hand side of Eq.~(\ref{drhoee}), with
$\rho_{\text{00}}$ being the population of Li-like ions, $j_{\text{xfel}}$ the photon flux of the XFEL pump pulse, and
$\sigma_{\text{0}}$ the K-shell photoionization cross section of the pump process ($1s^22l\rightarrow1s2l$). 
The XFEL pulse also depletes the $\left|\text{e}\right>$ and $\left|\text{g}\right>$ states, 
as modeled by the third term on the right-hand side of Eq.~(\ref{drhoee}) and the second term on the right-hand side of Eq.~(\ref{drhogg}), respectively, 
with $\sigma_{\text{e}}$ and $\sigma_{\text{g}}$ being the corresponding photoionization cross sections. 
In Eqs.~(\ref{drhoee})~and~(\ref{drhogg}), $\Gamma\rho_{\text{ee}}(x,t)$ describes spontaneous emission at rate $\Gamma$. 
In Eq.~(\ref{drhoeg}), the parameter 
\begin{eqnarray}
\gamma=\Gamma+\Delta\omega_{\text{e-i}}+(\sigma_{\text{e}}+\sigma_{\text{g}})j_{\text{xfel}}(x,t)
\label{gamma}
\end{eqnarray}
models the three contributions to the decay of the off-diagonal 
elements: $\Gamma$ is the decoherence originating from spontaneous photon emission;
the second term $\Delta\omega_{\text{e-i}}$ accounts for the broadening from electron--ion collisions~\cite{Griem1961S}; 
and the final term describes the contribution from depletion of the total population of He-like ions. 
$S(x,t)$ in Eq.~(\ref{drhoeg}) is a Gaussian white-noise term added phenomenologically, which satisfies $\left<S^*(x,t)S(z,t')\right>=F(z,t_1)\delta(t-t')$. 
For $E1$ transitions, one has~\cite{Larroche2000maxwellS}
\begin{eqnarray}
F(x,t)&=&\frac{\varepsilon_0\hbar\omega_0}{N_{\text{i}}\wp^2}\frac{d}{8\pi L}\frac{\gamma^2}{\omega_0^2}\Gamma \rho_{\text{ee}}(x,t),
\end{eqnarray}
where $\varepsilon_0$ is the vacuum permittivity and $N_{\text{i}}$ stands for the density of the ions in the plasma. 
$d=0.4$~$\mu$m is the radius of the XFEL spot on the plasma and $L$ is the length of the plasma. 

The coupling between the Maxwell equations and the Bloch equations is given through the induced fields $\mathcal{P}=-2N_{\text{i}}\wp\rho_{\text{eg}}$,
$\mathcal{M}=-2N_{\text{i}}m\rho_{\text{eg}}$ and $\mathcal{Q}=-2N_{\text{i}}q\rho_{\text{eg}}$ for $E1$, $M1$, and $M2$ transitions, respectively. 
Absorption of the XFEL pulse by the ions is included 
through the rate equations 
\begin{eqnarray}
\frac{\partial j_{\text{xfel}}}{\partial z}=-\sum_{\text{k}}\sigma_{\text{k}}\rho_{\text{kk}}N_{\text{i}}j_{\text{xfel}},
\label{djxfel}
\end{eqnarray}
where k=\{0, e, g\} represents the three different states. 

\subsection{Charge-state distribution} 
\begin{figure*}[ht]
\renewcommand{\figurename}{Figure}
\renewcommand{\thefigure}{S\arabic{figure}}
\includegraphics[width=0.7\textwidth]{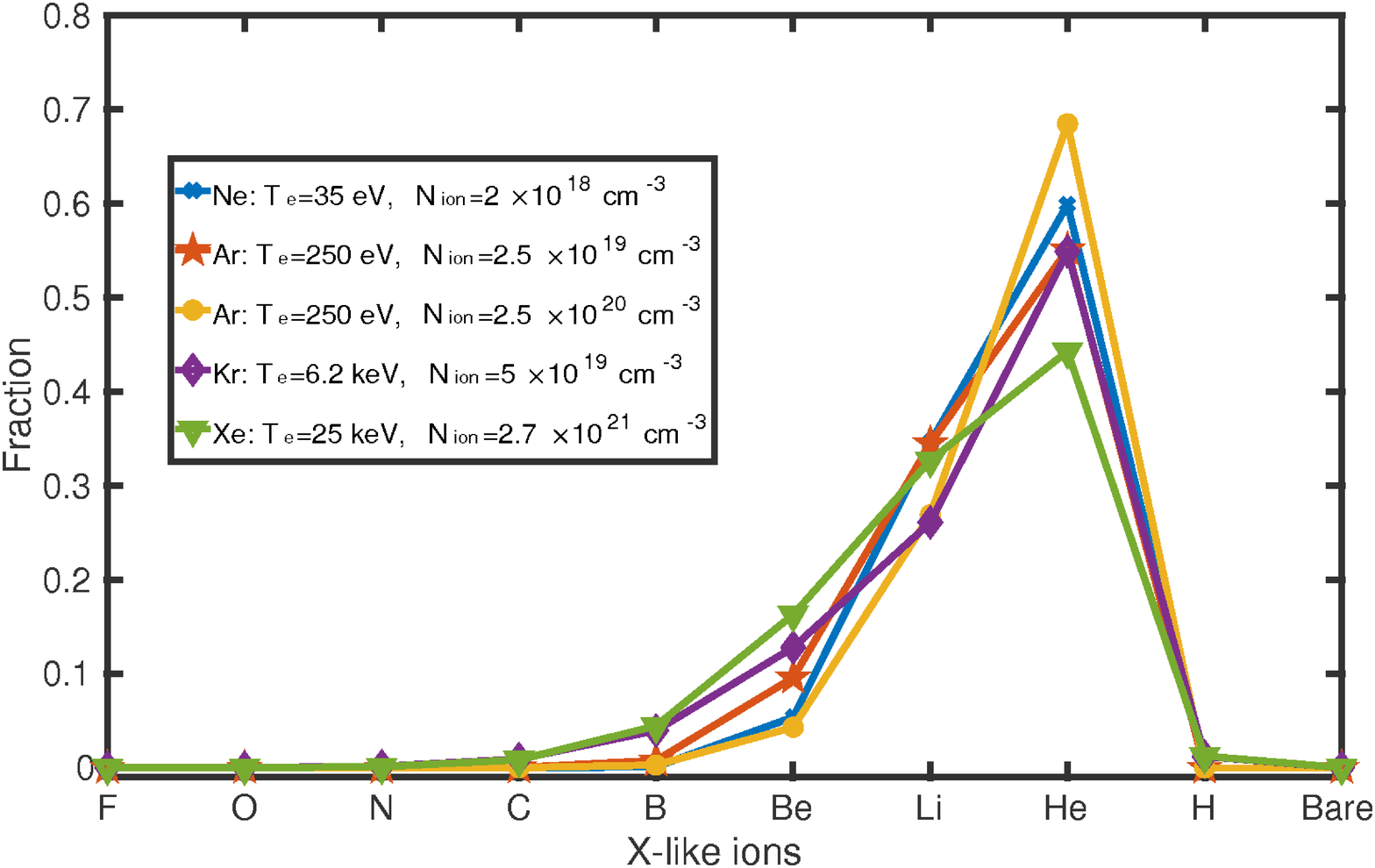}
\caption{\label{fig:CSD} \textbf{Charge-state distributions for different elements and simulation parameters.} }
\end{figure*}

Values of the initial populations of the states in Li- and He-like ions are computed with the FLYCHK code~\cite{Chung2005flychkS}
\begin{eqnarray}
\rho_{\text{00}}(x,t=0)&=&\rho_{_\text{Li-like}},\\
\rho_{\text{ee}}(x,t=0)&=&0,\\
\rho_{\text{gg}}(x,t=0)&=&\rho_{_\text{He-like}},
\end{eqnarray}
where $\rho_{_\text{Li-like}}$ and $\rho_{_\text{He-like}}$ are the fractions of Li-like ions ($1s^22l$ state) 
and He-like ions ($1s^2$ state) shown in Fig.~\ref{fig:CSD}. The initial population of the $1s2l$ upper lasing states in He-like ions is found to be negligible. 
Thus, most of the He-like ions are in the lower lasing state and no population inversion exists before XFEL-pulse pumping sets in. 
Since the decay of the upper lasing state in He-like ions is much slower than the K-shell photoionization rate, after the Li-like ions are pumped to a $1s2l$ state, 
this decays on a time scale much longer than the inverse of the K-shell photoionization rate. 
Population inversion develops after the lower lasing state ($1s^2$) of the He-like ions has been depleted by the XFEL pulse. 
In the simulation, only evolutions of the He-like ions are described by density-matrix theory. 
The evolution of the populations of the other charge states is modeled through rate equations. 

\subsection{XFEL parameters} 

Using the radiative parameters displayed in table~1 from the main text, and the corresponding K-shell and L-shell photoionization cross sections shown in table~\ref{Extended_data}, 
the XFEL parameters are fixed such that the requirements for lasing described in the main text and in Sec.~A1 are satisfied, 
and significant population inversion can be obtained. 
In particular, the XFEL photon energies $\omega_{\text{xfel}}$ listed in table~1 in the main text are tuned above the K-edge of the He-like ions, 
in order to ensure depletion of the initial population in the lower lasing state of He-like ions. For the elements considered, 
this photon energy is also above the K-edge of the corresponding Li-like ions. Photoionization of Li-like ions to a $1s2l$ excited state in 
He-like ions then ensures population inversion of the considered lasing transition.
Most of the XFEL bandwidths listed in table~\ref{Extended_data} are chosen taking into account realistic parameters at XFEL facilities in operation or under construction. 
Only for the case of $\text{Kr}^{\text{34+}}$ and $\text{Xe}^{\text{52+}}$, the smallest values of the XFEL coherence times, thus the largest values of the XFEL bandwidths, 
used in the simulations are limited by the time steps ($\delta t=0.0001\tau=0.95$~fs for $\text{Kr}^{\text{34+}}$ and $\delta t=0.001\tau=0.34$~fs for $\text{Xe}^{\text{52+}}$)
employed in our numerical calculations of the Maxwell--Bloch equations. 
However, simulations run with smaller time steps, thus larger XFEL bandwidths, provide results which are not significantly different. 
This is because photoionization pumping is determined by the XFEL flux and is not significantly influenced by the properties of the XFEL spectrum~\cite{Weninger2014S}. 
Therefore, when we compare the bandwidth of the XRLs and XFEL pulses in the main text, we refer to the realistic bandwidth measured at XFEL facilities and not to the values used in this table. 
\begin{table*}[ht]
\renewcommand{\tablename}{Table}
\renewcommand{\thetable}{S\arabic{table}}
\caption{\textbf{Extended XFEL parameters.} 
K-shell ionization cross section $\sigma_{\text{0}}$ and $\sigma_{\text{g}}$, together with the L-shell ionization cross section $\sigma_{\text{e}}$, are 
calculated from the LANL Atomic Physics Codes~\cite{LANL-CodesS} ($\text{kb}=10^{-21}\text{ cm}^2$). SASE XFEL parameters such as pulse duration, bandwidth and total photons inserted into the 
medium are simulated with a partial-coherence method~\cite{Cavaletto2012ResonanceS}. 
The number of photons absorbed during the lasing process is obtained from our numerical solutions of Eqs.~(\ref{drhoee}~--~\ref{djxfel}).}
\begin{ruledtabular}
\begin{tabular}{cccc|cccc}
\multicolumn{4}{c}{XFEL ionization cross section} & \multicolumn{4}{c}{XFEL parameters} \\
Upper state &  $  \sigma_{\text{0}}$  &  $\sigma_{\text{e}}$ &  $  \sigma_{\text{g}}$ & duration & bandwidth  & photons inserted & photons absorbed  \\
$1s2l$         & (kb) &           (kb)   & (kb)           &   (fs) & (eV)   &  &  \\
\hline
$\text{Ne}^{\text{8+}}~^1P_1   $ & $39.1 $ & $1.6 $   & $151.2 $  & 21.3  & $6.25$  & $2.4\times10^{12}$ & $6.3\times10^{9}$ \\
$\text{Ar}^{\text{16+}}~^3P_1  $ & $24.6 $ & $0.48 $  & $44.6  $  & 124   & $1.55$  & $2.9\times10^{12}$ & $1.0\times10^{12}$   \\
$\text{Ar}^{\text{16+}}~^1P_1  $ & $12.4 $ & $0.47 $  & $44.6 $  & 2.0   & $74$    & $3.8\times10^{12}$ & $2.7\times10^{10}$ \\
$\text{Kr}^{\text{34+}}~^3P_2  $ & $6.24 $ & $0.12$   & $10.9 $  & 202  & $0.57$ & $3.9\times10^{13}$ & $2.0\times10^{13}$   \\
$\text{Xe}^{\text{52+}}~^3P_2  $ & $2.66 $ & $0.06$   & $5.31 $  & 72.5  & $1.96$  & $5.6\times10^{13}$ & $3.6\times10^{13}$ \\
\end{tabular}
\label{Extended_data}
\end{ruledtabular}
\end{table*}
\newpage

\section{Line broadening in plasma}
In the following, we explain in detail the role of the presence of Doppler, collisional and Stark broadening effects, as they are also comparable to
the relatively small natural linewidth. As shown in table~1 from the main text, the Doppler broadening 
\begin{eqnarray}
\Delta\omega_{\text{D}}=\sqrt{\frac{8\text{ln}2k_{\text{B}}T_{\text{i}}}{m_{\text{i}}c^2}}\omega_{\text{0}},
\end{eqnarray} 
with $k_{\text{B}}$ being the Boltzmann constant, $T_{\text{i}}$ the ion temperature and $m_{\text{i}}$ the mass of the He-like ions, 
is not significant for light ions, but it becomes dominant for heavy ions like $\text{Kr}^{\text{34+}}$ and $\text{Xe}^{\text{52+}}$.

The electron-impact broadening is given by~\cite{Griem1961S}
\begin{eqnarray}
\Delta\omega_{\text{e-i}}&=&-\frac{ 16}{3\bar{v}_{\text{e}}} \frac{ N_{\text{e}}\hbar^2}{Z_{\text{i}}^2m_{\text{e}}^2} \text{ln}\Lambda~\left<\bm{r}\bm{r}\right>,
\end{eqnarray} 
with $\bar{v}_{\text{e}}=\sqrt{8k_{\text{B}}T_{\text{e}}/\pi m_{\text{e}}}$  being the average thermal velocity of the electrons in the plasma and $N_{\text{e}}$ the electron density, and 
$Z_{\text{i}}$ the charge number of the ions. $T_{\text{e}}$ and $m_{\text{e}}$ are the electron temperature and electron mass, respectively. 
$\text{ln}\Lambda\sim10$ is the Coulomb logarithm and $\bm{r}\bm{r}$ is a tensor with $\bm{r}$ being the dipole operator of the bound electrons in the ions. 
Such broadening is significant only for light ions such as $\text{Ne}^{\text{8+}}$ and $\text{Ar}^{\text{16+}}$ with lower electron temperatures, 
but becomes negligible compared to $\Delta\omega_{\text{D}}$ for heavy ions and higher electron temperatures. 

The quadratic Stark broadening from ion--ion interaction is calculated through~\cite{Griem1974S}
\begin{eqnarray}
\Delta\omega_{\text{i-i}}=\alpha\overline{F^2},
\end{eqnarray} 
with 
\begin{eqnarray}
\alpha&=&-\frac{1}{4\pi\hbar}\sum_{k\neq j}\wp_{jk}^2/(e E_{kj}^2), \\
\overline{F^2}&=&4Z_{\text{p}}^2e^2N_{\text{i}}^{\frac{4}{3}}/(\pi\varepsilon_0)^2,
\end{eqnarray} 
$\alpha$ is the quadratic Stark coefficient which would be different for each transition, 
and $\overline{F^2}$ is the mean-square electric-field strength generated by 
nearby perturbing ions with charge number $Z_{\text{p}}$. 
$\wp_{jk}$ and $E_{jk}$ are the electric-dipole moment and energy difference between the states $\left|\text{j}\right>$ and $\left|\text{k}\right>$, respectively. 
$\Delta\omega_{\text{i-i}}$, in general, is negligible for $N_{\text{i}}<10^{19}$ cm$^{-3}$, 
but becomes large for dense ion gases. 

In general, the natural broadening and electron--ion impact broadening are homogeneous for each ion, giving a Lorentzian spectrum. 
The Doppler broadening and ion--ion Stark broadening, on the other hand, are inhomogeneous for different ions, and result in a Gaussian spectrum. 
For systems involving both homogeneous and inhomogeneous broadenings, as it is the case here, 
the real spectrum has a Voigt line shape given by the convolution of the Lorentzian and Gaussian profiles, 
with a FWHM $\Delta\omega_{\text{V}}$ approximately given by~\cite{Olivero1977EmpiricalS} 
\begin{eqnarray}
\Delta\omega_{\text{V}}&=&0.5364\Delta\omega_{\text{L}}+\sqrt{0.2166\Delta\omega_{\text{L}}^2+\Delta\omega_{\text{G}}^2}.
\label{Voigt}
\end{eqnarray} 
Here, $\Delta\omega_{\text{L}}=\Gamma+\Delta\omega_{\text{e-i}}$ is the FWHM of the Lorentzian function, and $\Delta\omega_{\text{G}}=\Delta\omega_{\text{D}}+\Delta\omega_{\text{i-i}}$ is the 
FWHM of the Gaussian function. 

Numerical simulations of the lasing process accounting for the inhomogeneous broadening should take into account the distributions of the thermal velocity as well as 
Stark shift of the ions, which renders the simulations time consuming. 
However, Eq.~(\ref{Voigt}) shows that 
\begin{eqnarray}
\Delta\omega_{\text{V}}\sim\Delta\omega_{\text{L}}+\Delta\omega_{\text{G}}=\Gamma+\Delta\omega_{\text{e-i}}+\Delta\omega_{\text{D}}+\Delta\omega_{\text{i-i}}
\end{eqnarray} 
For the sake of simplicity, we can approximate the parameter in Eq.~(\ref{gamma}) as 
\begin{eqnarray}
\gamma=\Gamma+\Delta\omega_{\text{e-i}}+\Delta\omega_{\text{D}}+\Delta\omega_{\text{i-i}}+(\sigma_{\text{e}}+\sigma_{\text{g}})j_{\text{xfel}}(x,t).
\end{eqnarray} 
With this approximation, the distribution of the ions over different thermal velocities and Stark shifts is automatically included in the Maxwell--Bloch equations. 
This simplification may lead to a maximum of 25\% overestimate of the bandwidth compared to the Voigt bandwidth $\Delta\omega_{\text{V}}$ in the simulations. 
However, this will not change our conclusions in the main text.

\section{Other simulation results} 
\begin{figure*}[ht]
\renewcommand{\figurename}{Figure}
\renewcommand{\thefigure}{S\arabic{figure}}
\includegraphics[width=0.95\textwidth]{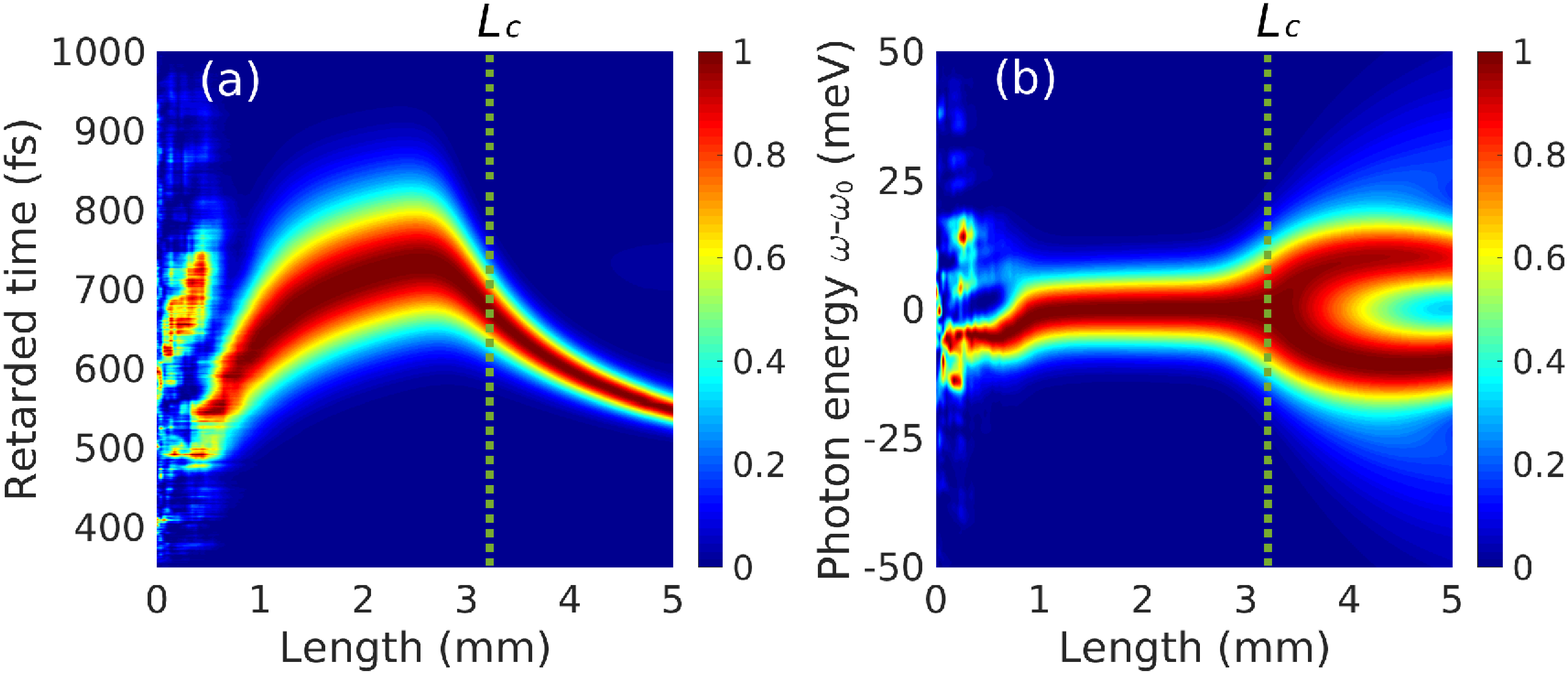}
\caption{\label{fig:shot2} \textbf{Evolution of the normalized x-ray laser intensity and spectrum (shot~2).} 
Results for SASE-pulse shot~2 in Fig.~3f in the main text; green dotted lines in Figs.~3c,d in the main text. 
\textbf{a}, Intensity shown as a function of retarded time and propagation length. 
\textbf{b}, Power spectrum displayed as a function of photon energy and propagation length. 
For a given length, the intensity and spectrum are normalized to the maximum value of the corresponding 
profiles at such length. The vertical dotted lines indicate the characteristic length $L_{\text{c}}$ defined in the main text. 
The decrease in the XRL bandwidth at the end of the medium is a result of XFEL absorption.}
\end{figure*}

\begin{figure*}[ht]
\renewcommand{\figurename}{Figure}
\renewcommand{\thefigure}{S\arabic{figure}}
\includegraphics[width=0.95\textwidth]{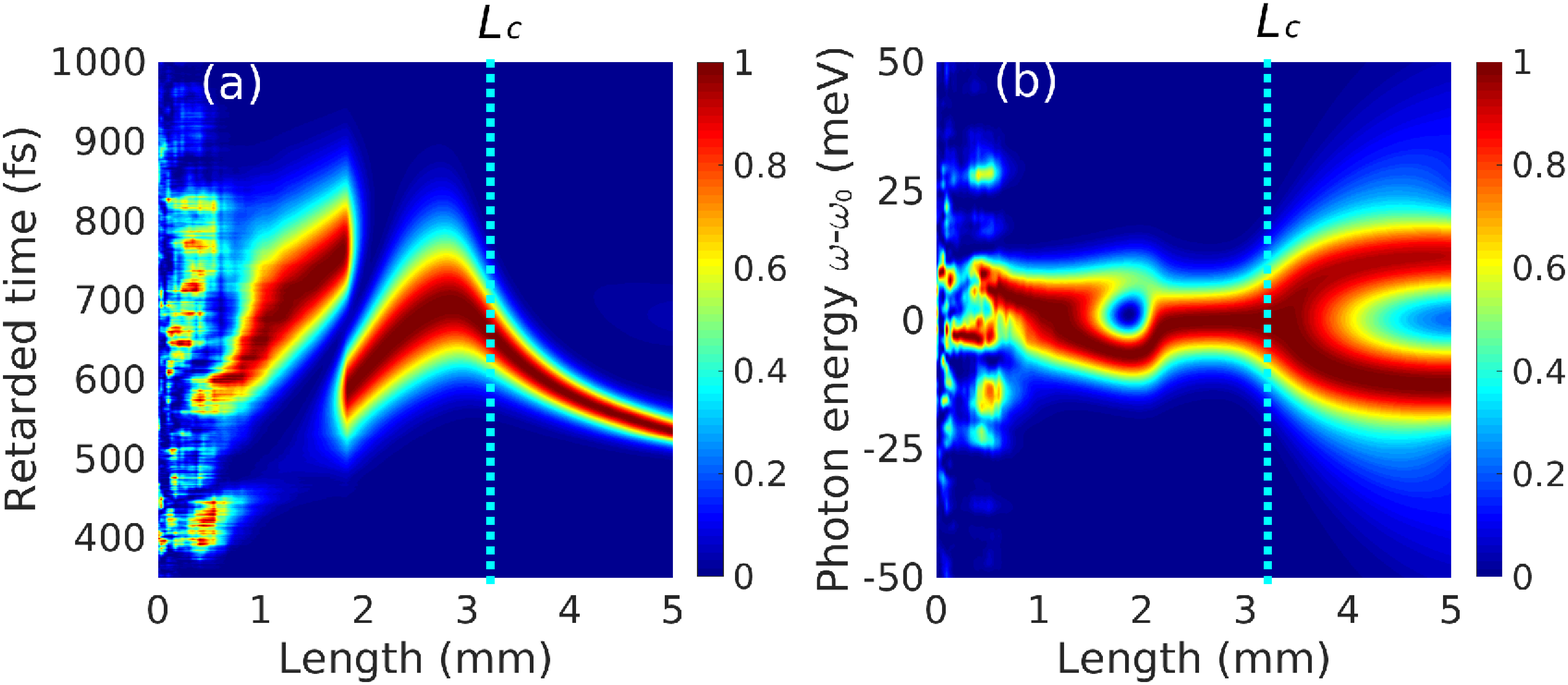}
\caption{\label{fig:shot3} \textbf{Same as Fig.~\ref{fig:shot2} for shot~3.} 
Results for SASE-pulse shot~3 in Fig.~3g in the main text; blue dotted lines in Figs.~3c,d in the main text. 
The exotic structures in both intensity and spectrum around $L=2$ mm originate from 
XFEL photoionization of the upper lasing state. The decay of this state is different during and after the XFEL pulse. This renders it possible that two peaks develop 
and propagate before saturation.}
\end{figure*}

\begin{figure*}[ht]
\renewcommand{\figurename}{Figure}
\renewcommand{\thefigure}{S\arabic{figure}}
\includegraphics[width=0.95\textwidth]{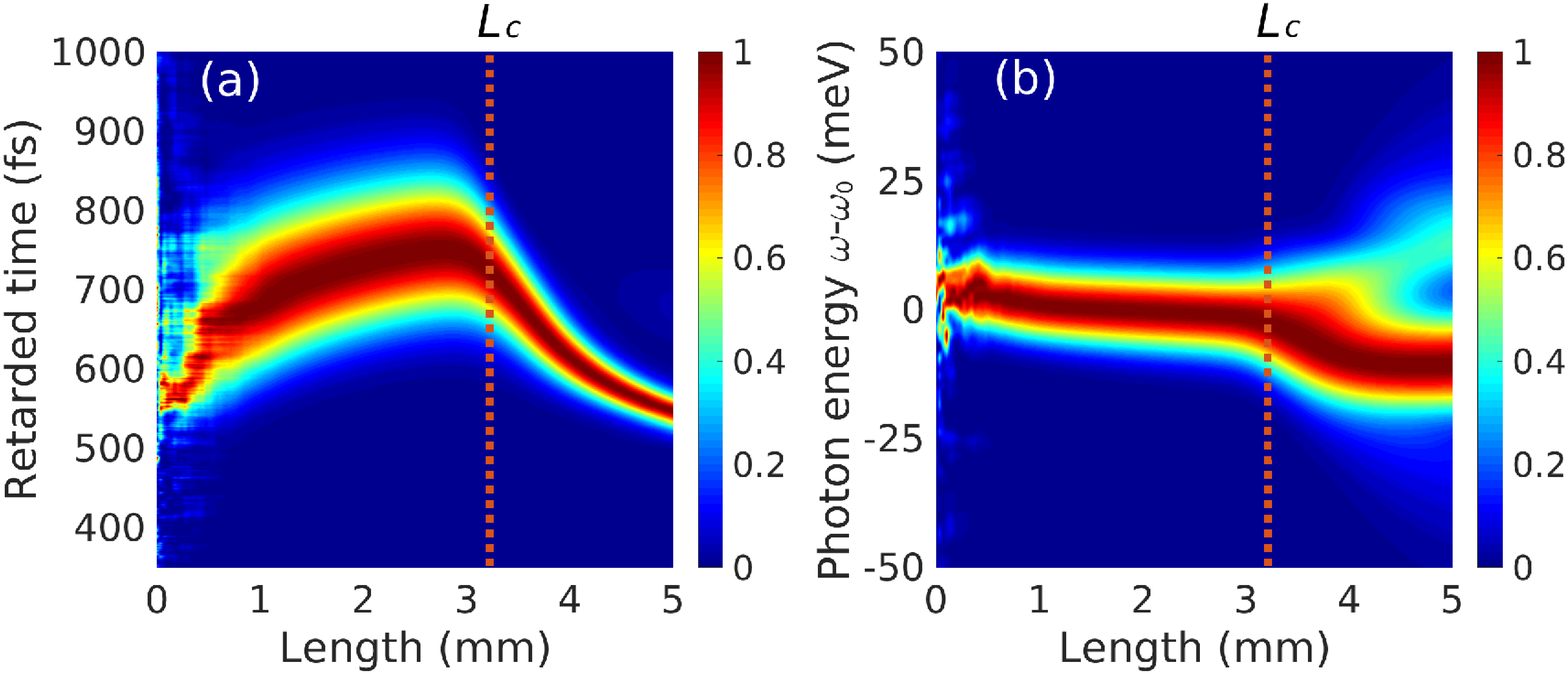} 
\caption{\label{fig:shot4} \textbf{Same as Fig.~\ref{fig:shot2} for shot~4.} 
Results for SASE-pulse shot~4 in Fig.~3h in the main text; red dotted lines in Figs.~3c,d in the main text. 
A slight shift of the peak of the spectrum from $\omega_0$ is here apparent.}
\end{figure*}

\begin{figure*}[ht]
\renewcommand{\figurename}{Figure}
\renewcommand{\thefigure}{S\arabic{figure}}
\includegraphics[width=0.95\textwidth]{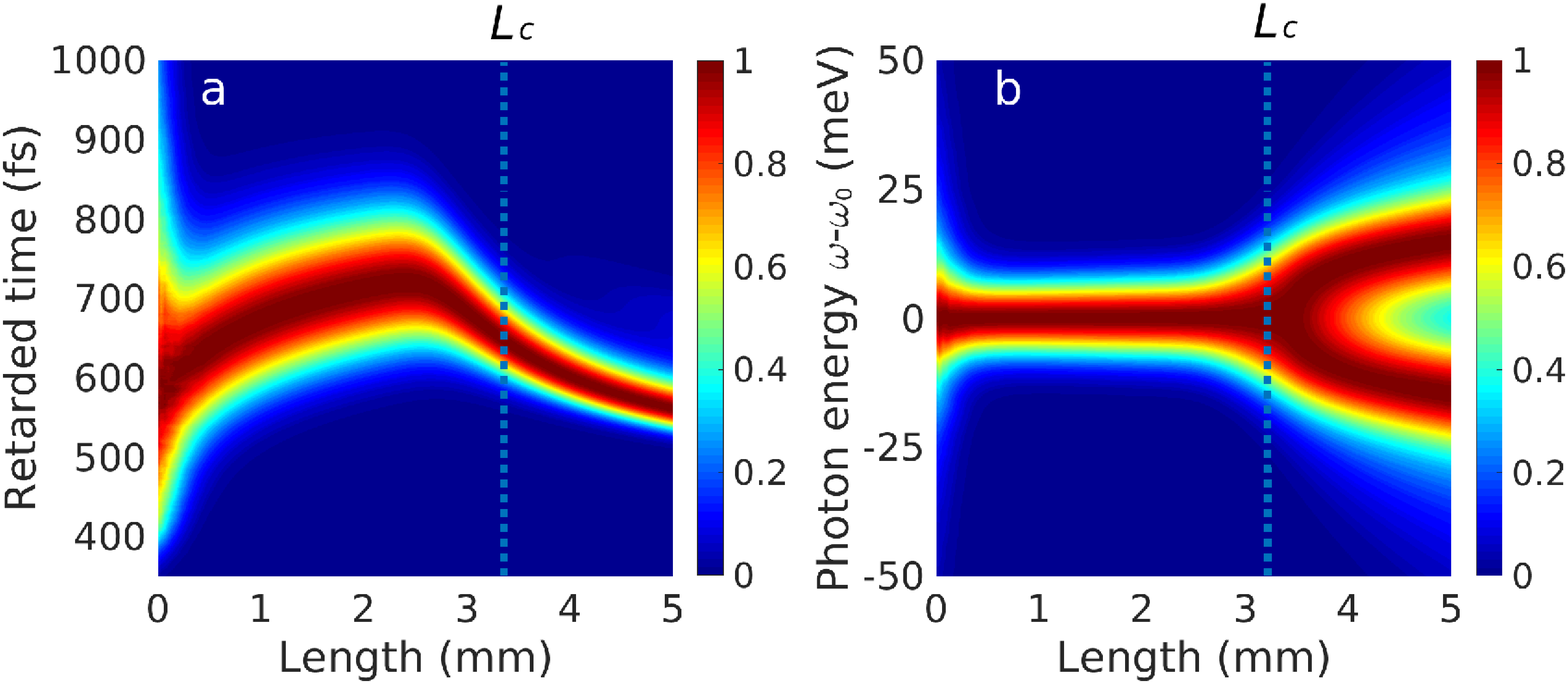}
\caption{\label{fig:average} \textbf{Evolution of the normalized x-ray laser intensity and spectrum averaged over 1,000 realizations of SASE XFEL pulses.} 
Intensity and spectra evaluated at the characteristic length $L_{\text{c}}$ are exhibited by the solid lines in Figs.~3c,d in the main text.
The chaotic features for small propagation lengths, which are shown for single simulations in Figs.~3a,b in the main text and in Figs.~\ref{fig:shot2}--\ref{fig:shot4} above, become smooth here.}
\end{figure*}

Simulation results of the intensity profile and spectrum as a function of propagation length for shots~2--4 in Fig.~3 in the main text, 
as well as the intensity profile and spectrum averaged over 1,000 simulations, are displayed in Figs.~\ref{fig:shot2}--\ref{fig:shot4} and Fig.~\ref{fig:average}, respectively. 
Shot-to-shot differences are a consequence of the random SASE-XFEL-pulse profiles used, shown in Figs.~3e-f in the main text. 

\newpage

\end{document}